# High-performance 2D 1×N T-junction Wavelength (De)Multiplexer Systems by Inverse Design

Yusuf A. Yilmaz[1,†,*], Ahmet M. Alpkiliç[1,†], Aydan Yeltik[2] and Hamza Kurt[1]

[1] Department of Electrical and Electronics Engineering, TOBB University of Economics and Technology, Ankara 06560, Turkey
[2] Department of Physics, University of Cambridge, Cambridge CB3 0HE, UK
*Corresponding author: yayilmaz@etu.edu.tr
† Y. A. Yilmaz and A. M. Alpkilic contributed equally.

Previously proposed designs of integrated photonic devices have used the intuitive brute force approach or optimization methods that employ parameter search algorithms. However, a small parameter space and poor exploitation of the underlying physics have limited device performance, functionality, and footprint. In this paper, we propose efficient and compact 1×N in-plane-incidence wavelength demultiplexers by using recently developed objective-first inverse design algorithm. Output ports in the presented 1×N photonic devices are located along the transverse to the input channel. Ultra-high device performance was achieved for the specific designs of 1×2, 1×4, and 1×6 wavelength (de)multiplexers with small footprints 2.80 μm x 2.80 μm, 2.80 μm x 4.60 μm, 2.80 μm x 6.95 μm, respectively. We used two approaches to binarization—level-set and binarization-cost—to obtain silicon wavelength demultiplexer considering fabrication constraints. For instance, the transmission efficiency of binarization-cost 1×2 demultiplexer was -0.30 dB for 1.31 μm and -0.54 dB at 1.55 μm while crosstalk at the operating wavelengths are negligibly small, i.e., -17.80 and -15.29 dB, respectively. Moreover, for the binarization-cost 1×4 demultiplexer, the transmission efficiency values were approximately -1.90 dB at 1.31, 1.39, 1.47, and 1.55 μm as the crosstalk was approximately -13 dB. Furthermore, the objective-first algorithm was used to employ our demultiplexers as multiplexers which means the ports that were once used as inputs in demultiplexers are designed to be used as outputs. The inverse design approach that allows for the implementation of more than six output channels together with the proposed functionalities can help develop compact and manufacturable 1×N couplers.

**OCIS codes:** *(000.4430) Numerical approximation and analysis; (060.4230) Multiplexing; Inverse design; Objective-first algorithm; Wavelength (de)multiplexing.*

1. **INTRODUCTION**

Passive optical network technologies have attracted a great deal of interest owing to their significant potential for various applications, including high-speed, broadband optical communication, and effective optical sensing systems [1,2]. Therefore, limitations on the capacity of these unpowered optical systems have been widely investigated using several low-energy-dissipating and straightforward routes, particularly to increase transmission rates and the density of up/down-stream bandwidth [3,4]. For this purpose, different multiplexing structures, those featuring divisions of time, mode, and wavelength have been proposed to date [5–19]. Nanophotonic designs have shown to have considerably efficient structures that can be implemented to obtain favorable characteristics, such as low loss, high sensitivity, and clear index contrast in dielectric distribution [7,8,10,18]. Furthermore, the integration of such all-dielectric nanophotonic components as multiplexing couplers, waveguides, lasers and LEDs into a single device may lead to the miniaturization of optical circuits with a high data processing capability, as in silicon chips used for integrated electronics. However, contrary to electronic circuits obtained through various well-defined design languages, there is a lack of effective design methodologies in nanophotonics [20,21].

Nanophotonic designs have been generally obtained using intuitive hand-tuning approaches with an analytical basis or various optimization methods via parameter search algorithms [10,14,22–30]. Of them, all-dielectric photonic-based designs of wavelength demultiplexers (WDMs) may boast numerous important features, including high transmission efficiency, high spectral resolution, low crosstalk, cost-effectiveness, and compactness [31,32]. Photonic crystal (PC)-based WDMs possessing different architectures, e.g., T-shaped and P-shaped structures, have been designed mostly through heuristic/analytical methods [11,12,33,34]. These all-dielectric unpowered optical splitters have been designed as coarse or dense multi-channel devices, depending on needs of the application in terms of channel capacity, transmission distance, and operating region along the electromagnetic spectrum. Parameter search optimization has also been used to the design photonic-based WDMs [13,35,36]. A wide variety of photonic WDM designs highlights the importance of these devices as promising candidates for easy implementation to newly emerging on- and off-chip systems in optical communication networks [37,38]. However, a small parameter space and the poor exploitation of the underlying physics in the design methodologies limits device performances in terms of functionality, transmission efficiency, and footprint. Recently, a highly promising alternative approach called the objective-first inverse design algorithm was proposed, and various integrated photonic components have been shown to deliver high performance using this cost-effective technique [14,15,39–42]. The performance metrics ($\vec{E}, \vec{H}, \vec{S}$) of nanophotonic devices are first determined as the target parameters defined in the cost function and the dielectric distribution ($\varepsilon$) of these structures is obtained in an iterative manner.

Promising all-dielectric WDM structures designed using the objective-first algorithm with different architectures have been proposed in a number of studies [14–16,43]. A report released in 2014 presented the first fabricable design of a highly efficient vertical-incidence wavelength-demultiplexing grating coupler [14]. In another study by Vuckovic et al., a compact broadband WDM containing 1×2 parallel in/out waveguide channels were demonstrated numerically and experimentally [15]. The results of simulations and experiments on a 2.8 μm × 2.8 μm WDM showed a transmission of approximately -2 dB at target wavelengths of 1.3 and 1.55 μm. The inverse design of a 1×3 fabricable WDM with a channel spacing of 40 nm was reported in 2017 via the objective-first methodology by imposing curvature constraints on the device boundaries [40]. The design showed a simulated transmission contrast higher than 16 dB at the channel centers in the range of 1.49-1.57 μm, and with an insertion loss of ~1.5 dB. Another design by the same group using the same algorithm provided a compact (4.5 μm × 5.5 μm) narrowband WDM with three outdrop channels at a 40 nm spacing [16]. As design performance metrics, relatively high transmission efficiencies were obtained at around -1.5 dB (simulation) at wavelengths of 1.5, 1.54, and 1.58 μm, and of around -2.5 dB (experiment) at shifting peak wavelengths, with crosstalk under -15 dB and -10.7 dB, respectively. To achieve fabricable WDMs, the objective-first algorithm has been combined with various design techniques in the literature, e.g., steepest descent and biasing [15,16]. In a recent work, a similar inverse design approach combined with a gradient-based algorithm has been proposed to the form wavelength demultiplexing grating couplers with different geometries, such as a pass-through architecture, exhibiting reasonably high efficiencies at such specified wavelengths as 1.31 and 1.49 μm [43]. It can be inferred from these efforts that the proposed inverse design algorithm is highly suitable for next-generation, compact, and manufacturable nanophotonic devices with novel functionalities and features, including high-efficiency transmission and lower sensitivity to fabrication errors.

In this study, we propose and demonstrate highly efficient and ultra-compact 1×N in-plane-incidence T-junction wavelength (de)multiplexing systems. The designs presented here are 1×2, 1×4 and 1×6 WDMs. All-dielectric WDM structures designed with short simulations provide near-unity transmissions, a small footprint, significantly low crosstalk, and channel spacings as small as 80 nm. We use the objective-first inverse design algorithm as in the aforementioned examples. The difference between this study and past relevant work is that we accomplish the selection of the largest wavelength for up to six channels as well as the simultaneous inverse operation of the structures as multiplexers. In addition to the proposed design method based on the finite-difference frequency-domain (FDFD) approach, the finite-difference time-domain (FDTD) method is used to confirm transmission efficiency at the specified wavelengths and obtain the spectral transmissions of the proposed WDMs [44].

## 2. OPTIMIZATION AND BINARIZATION METHODS

The inverse-design methodology is applied in two steps: continuous optimization and binarization. Both $\varepsilon^{-1}$ and $\vec{H}$ are bi-linear in the following electromagnetic wave equation:

$$\nabla x \varepsilon^{-1} \nabla x \vec{H} - \mu_0 \omega^2 \vec{H} = \nabla x \varepsilon^{-1} \vec{H}, \qquad (1)$$

where $\varepsilon$ is the space-dependent value of the material dielectric, $\vec{H}$ is the magnetic field, $\mu_0$ is permeability in vacuum, $\omega$ is angular frequency, and $\vec{J}$ is the excitation current density. By formulating the problem in the language of linear algebra and applying the necessary replacements, the fields are forced to satisfy the performance objectives first and the physical violations are minimized by attempting to satisfy Maxwell's equations [45]. In this way, structure is obtained by alternatingly changing between the two sub-problems, which are $\min_{\vec{H}} \|A(\varepsilon^{-1})\vec{H} - b(\varepsilon^{-1})\|^2$ subject to $f(\vec{H}) = f_{ideal}$, and $\min_{\varepsilon^{-1}}\|C(\vec{H})\varepsilon^{-1} - d(\vec{H})\|^2$ subject to $\varepsilon_{min}^{-1} \leq \varepsilon^{-1} \leq \varepsilon_{max}^{-1}$. Herein, $f_{ideal}$ and $\|\ \|^2$ are the desired electromagnetic behavior function and the Euclidean operator, respectively. The convergence of the algorithm to the desired structure is performed through convex optimization using the CVX MATLAB package [46]. The final structure is obtained as a continuous dielectric distribution. In this study, the dielectric limits were defined as $\varepsilon_{air} \leq \varepsilon \leq \varepsilon_{si}$ for continuous dielectric design.

In practice, photonic devices have a discrete dielectric distribution through a set of materials, such as binary structures composed of only silicon and air. Different approaches can be applied to this binarization process. The simplest method is the level-set method, in which the values higher than a specified threshold ($\varepsilon_{th}$) converge to the dielectric value of silicon while smaller values approach the value of air. Mathematically, a threshold dielectric constant is specified to binarize the structure as $\varepsilon = \varepsilon_{Si}$ if $\varepsilon > \varepsilon_{th}$ and $\varepsilon = \varepsilon_{air}$ if $\varepsilon \leq \varepsilon_{th}$. The choice of $\varepsilon_{th}$ is significant for the target performance criteria and needs to be determined for each design. Moreover, the efficiency of the level-set method is strongly dependent on the dielectric distribution of the continuous structure. İf the structure is concentrated in the intermediate values instead of the final values of the determined range of dielectric constant, it is challenging to obtain useful results via the level-set method. Therefore, the binarization-cost method contains a cost to extract a binary design from the continuous distribution to be used [42]. The value of $\varepsilon_{th}$ is determined by searching the minimum of the physical residual obtained for the binary structure. Following this, binarization-cost is included into $\min_{\varepsilon^{-1}}\|C(\vec{H})\varepsilon^{-1} - d(\vec{H})\|^2$ as in the following expression:

$$\min_{\varepsilon^{-1}}\|C(\vec{H})\varepsilon^{-1} - d(\vec{H}) + \gamma(\varepsilon^{-1} - \varepsilon_{bin}^{-1})\|^2, \qquad (2)$$

where $\varepsilon_{bin}^{-1}$ and γ are the binary values at in the previous iteration and the weighting factor, respectively. In the binarization-cost method, γ limits the extent to which the new proposed distribution, $\varepsilon^{-1}$, can be removed from the discrete distribution in the previous iteration $\varepsilon_{bin}^{-1}$. In the same study, *B*, which shows how discrete the structure is, is defined [42]. It is equal to one when the structure is completely binary and is zero when the structure is completely the average of the dielectric constant of silicon and air. The *B* parameter of the structure produced in Eq. (2) is greater in value than in the algorithm without the addition of binarization-cost, but the resulting structures are not completely binary. Therefore, to obtain the discrete structure, the level-set approach should be applied. If the *B* parameter of the structure produced by the algorithm is close to one, there are acceptable variations in the performance criteria when the structure is completely binarized by the level-set method.

For the success of the binarization-cost method, γ and the number of iterations should be chosen appropriately depending on the number of wavelengths, dimensions and physical residue. The value of γ must be slightly increasing from zero to ∞ as the smallest value of the binarization-cost term in Eq. (2) allows for the desired behavior of the magnetic field to occur in the initial iterations. In subsequent iterations, the binarization effect becomes more important with increasing γ. The desired functionality of the WDMs in this study, called as "performance objective", is transmission efficiency at the defined peak wavelengths. The increase in γ, depending on the number of iterations, reduces the effect of the performance objective. For this reason, the mathematical form and speed of an increase in γ varies by design and should be investigated separately for each design.

### 3. DEMONSTRATIONS OF 1×N T-JUNCTION (DE)MULTIPLEXING DESIGNS

A schematic representation of the proposed 1×N all-dielectric WDMs is shown in Fig. 1. The output waveguides in the structures are located along the transverse to the input waveguide channel. All WDMs are designed for a light source with a fundamental TE polarization (with non-zero components of $\vec{E}_x$, $\vec{E}_y$, and $\vec{H}_z$). The target wavelengths are associated with port numbers such that the shortest wavelength is directed into $P_1$ while the longest wavelength is directed to $P_N$. The wavelengths are defined by considering the telecommunication bands and the lattice constant, $a$, was determined to be 37 nm. Our 1×N design approach is also applicable to wavelength tunable T-junction WDMs depending on the needs of the application. The width of the waveguides was specified as 17$a$ (629 nm). The initial dielectric constant of all continuous structures was set to the permittivity of silicon (12.25). The aforementioned level-set and binarization-cost methods were used for the discretization of the continuous structures. The workstation used in this study was the Intel(R) Xeon(R) CPU E5-2650 v4 (x2) with 96 GB of RAM.

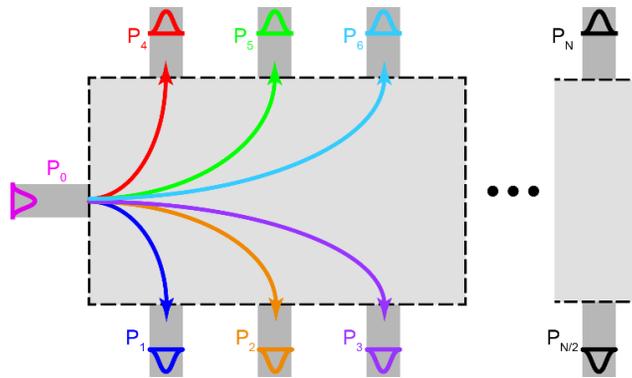

Fig. 1. Schematic representation of the proposed 1×N all-dielectric optical WDMs.

Fig. 2 shows the inverse-designed 1×2 ultra-compact photonic demultiplexing devices with a footprint of 2.80 μm × 2.80 μm (76$a$ x 76$a$). The device in Fig. 2(a) with a continuous dielectric distribution was obtained in ~6 h, a significantly short time of simulation compared with that in the related literature [5]. As shown in the figure, the WDM splits the incident mode-like broadband light in the range of 1.2—1.7 μm into beams of two specified targets $\lambda$, which are 1.31 μm (O-band) as transmitted through the lower waveguide and 1.55 μm (C-band) through the upper waveguide. The transmission spectra of the device in Fig. 2(a) are presented in Fig. 2(b). The transmission efficiency was -0.06 dB at 1.31 μm and -0.70 dB at 1.55 μm, highly promising at higher than 85% efficient. Crosstalk at the operating wavelengths was negligibly small, i.e., -26.01 and -13.88 dB, and was calculated at the corresponding output ports. Furthermore, Fig. 2(c) shows the binary structure obtained via the level-set approach in Fig. 2(a) structure with $\varepsilon_{th} = 4.5$. The optimal $\varepsilon_{th}$ is calculated numerically so that physical residual is minimized. The transmission efficiencies of the level-set binary structure shown in Fig. 2(d) were -0.59 dB at 1.31 μm and -0.95 dB at 1.55 μm, in addition to crosstalk at the output ports of -13.58 and -12.78 dB, respectively. Compared with the continuous and the level-set designs, the reduction in transmission efficiency is acceptable because the initial dielectric constant of the continuous design is the value of the dielectric constant of silicon. When the structure in Fig. 2(a) was examined, it consisted mainly of silicon, i.e., the intermediate dielectric values were low. Fig. 2(e) shows the binary structure obtained by using the binarization-cost method. $\varepsilon_{th}$ is 6.0 while γ is specified as a constant $10^{-12}$ in each iteration. The transmission efficiencies were -0.30 and -0.54 dB in increasing order of targeted wavelength, and crosstalk was -17.80 dB for 1.31 μm and -15.29 dB for 1.55 μm. The number of iterations was increased to create a structure that provided acceptable performance criteria. The number of iterations was increased because the algorithm performed discretization in addition to providing performance criteria for two wavelengths. Referring to the normalized transmission values of Figs. 2(d) and (f), it is clear that in addition to an increase in the transmission value at a wavelength of 1.55 μm, the unwanted peak around 1.40 μm was removed.

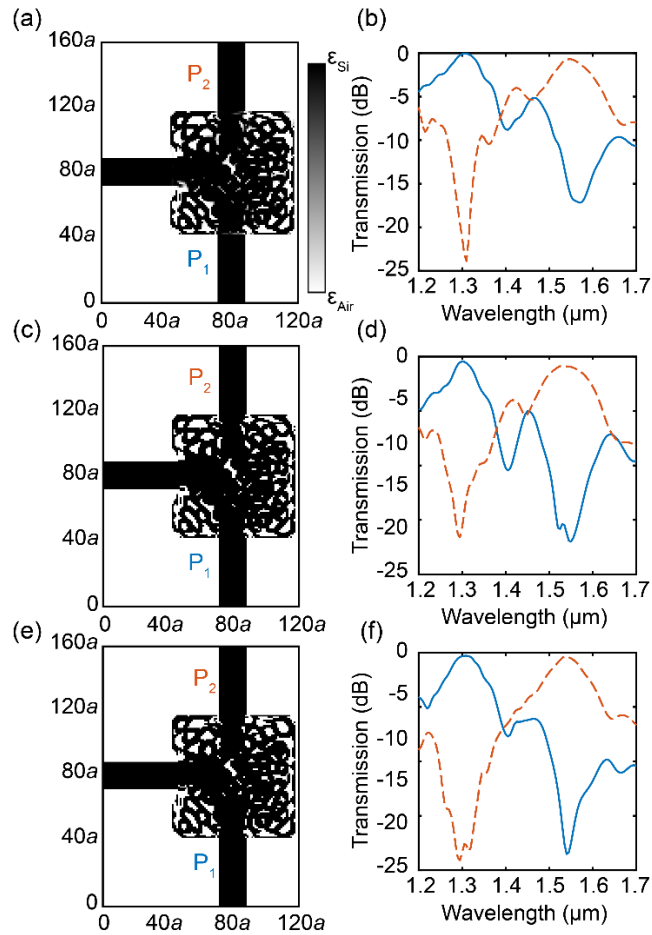

Fig. 2. (a) All-dielectric 1×2 T-junction WDM with continuous distribution, (b) normalized transmission efficiency of (a), (c) WDM with binary distribution obtained through the level-set method, (d) normalized transmission efficiency of (c), (e) WDM with binary distribution obtained through the binarization-cost method and (f) normalized transmission efficiency of (e).

Other WDM designs with a channel configuration of 1×4 are presented in Fig. 3. The designs are elongated along the propagation direction of the added channels, resulting in a footprint of 2.80 μm × 4.60 μm (76a x 124a). A continuous distribution of 1×4 T-junction WDM is shown in Fig. 3(a), where the simulation lasted for ~18 h. Fig. 3(b) shows transmissions in the range [1.20—1.70] μm, where the design possess ultimately yields high transmission efficiencies at the specified wavelengths: -0.57 dB at 1.31 µm, -0.71 dB at 1.39 µm (E-band), -1.25 dB at 1.47 µm (S-band), and -2.11 dB at 1.55 µm. In addition to the high transmission performance, crosstalk values of the device were in the range [-13.05—-16.66] dB. The dielectric distribution of the binary structure resulting from the level-set method for the structure in Fig. 3(a) with a numerically determined $\varepsilon_{th} = 4.3$ is shown in Fig. 3(c), and the normalized transmission efficiencies are given in Fig. 3(d). The transmission efficiencies of the level-set binary structure presented in Fig. 3(d) are -0.92 dB at 1.31 µm, -1.65 dB at 1.39 µm, -2.07 dB at 1.47 µm, and -2.23 dB at 1.55 µm. As the $B$ value of the continuous structure in Fig. 3(a) is 0.95, there is an acceptable decrease in transmission values when the structure is binarized using the level-set method. The binary structure in Fig. 3(e) is obtained using the binarization-cost method, with $\varepsilon_{th} = 6$ and $\gamma = 10^{-12}$*iteration number. In Fig. 3(f), the transmission efficiencies of the binary structure of the binarization-cost were -1.86 dB at 1.31 µm, -1.89 dB at 1.39 µm, -1.77 dB at 1.47 µm, and -1.87 dB at 1.55 µm and crosstalk was around ~-13 dB. Compared with the 1×2 design, difference in transmission efficiency between the continuous and binary structures was high because of more objectives in 1×4 design. In the binarization-cost method, as the sum of binarization cost and the physical residuals of each frequency value were minimized, the effect of binarization decreased with increasing frequency.

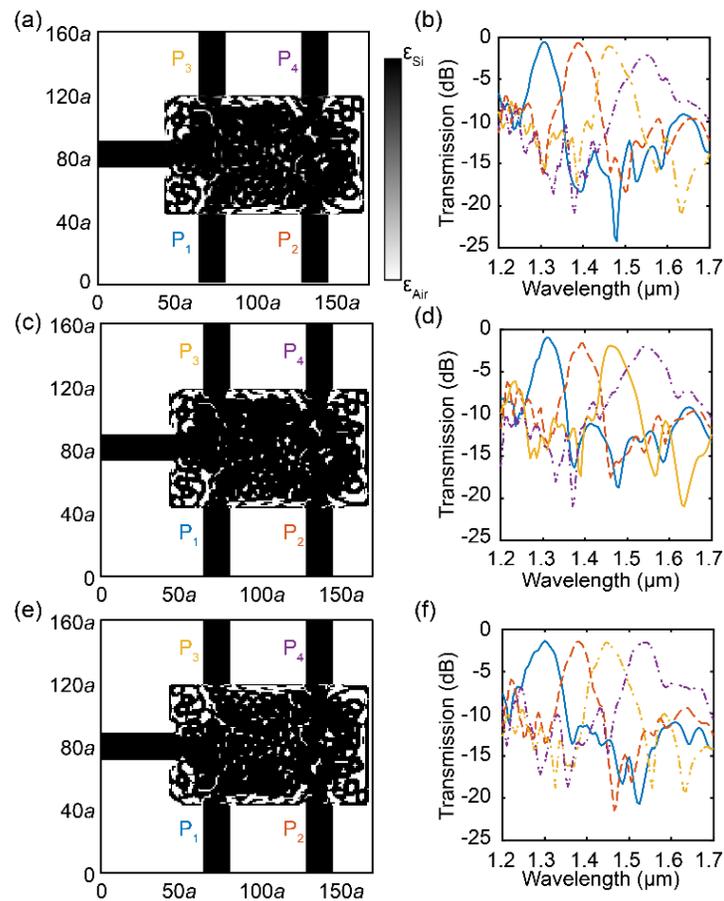

Fig. 3. (a) All-dielectric 1×4 T-junction WDM with continuous distribution, (b) normalized transmission efficiency of (a), (c) WDM with binary distribution obtained through the level-set method, (d) normalized transmission efficiency of (c), (e) WDM with binary distribution obtained through the binarization-cost method and (f) normalized transmission efficiency of (e).

For further confirmation of the design approach used in this study, we tested 1×6 WDMs with small device dimensions of 2.80 μm × 6.95 μm (76*a* × 192*a*), as shown in Fig. 4. To the best of our knowledge, this is the first demonstration of an inversely designed WDM with six output ports. The continuous device shown in Fig. 4(a) was designed in approximately 110 h. Owing to the complexity of the structure, the simulation time significantly increased compared with that for the other structures. Transmissions spanning wavelength spectra of [1.20–1.70] μm are given in Fig. 4(b). As in the other structures of the proposed 1×N demultiplexing system, the 1×6 T-junction WDM device delivered ultra-high transmissions at the defined wavelengths, which were -0.31 dB at 1.23 μm (shorter wavelength), -0.90 dB at 1.31 μm, -0.70 dB at 1.39 μm, -1.14 dB at 1.47 μm, -1.80 dB at 1.55 μm, and -2.12 dB at 1.63 μm (U-band). The device also possessed low channel spacing, 80 nm, with potential for higher channel resolution. Significantly low crosstalk was also observed in the designed structure, at -25.63, -20.23, -21.72, -17.72, -16.63, and -15.86 dB for output waveguides in increasing order of wavelength. The level-set approach with $\varepsilon_{th} = 4.3$ is shown in Fig. 4(c). In Fig. 4(d), the transmission efficiencies of the structure generating the level-set method are -1.33 dB at 1.23 μm, -2.13 dB at 1.31 μm, and -1.32 dB at 1.39 μm, -1.19 dB at 1.47 μm, -2.25 dB at 1.55 μm, and -2.20 dB at 1.63 μm; and crosstalk was -17.92, -15,91, -18.42, -18.51, -15.39, and -15.51 dB, respectively. These acceptable transmission efficiency values indicate that the algorithm successfully designed the T-junction WDMs although the wavelength increased. The performance of the binarization-cost approach decreased with increasing wavelength ports. For a two-wavelength design, it is better than the level-set method: With increasing number of wavelengths, the efficiency of the level-set method is higher than that of the binarization-cost method. Therefore, the binarization-cost approach was not applied to 1×6 design.

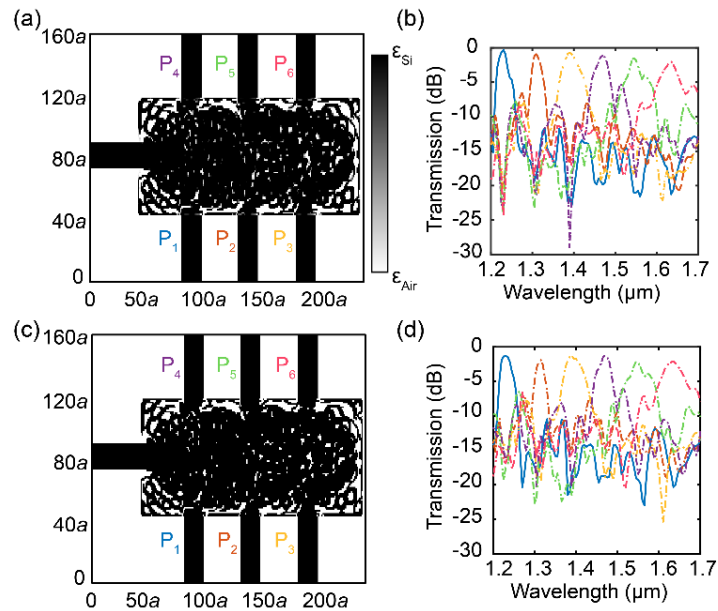

Fig. 4. (a) All-dielectric 1×6 T-junction WDM with continuous distribution, (b) normalized transmission efficiency of (a), (c) WDM with binary distribution obtained through the level-set method and (d) normalized transmission efficiency of (c).

The CVX optimization package encouraged us to utilize our demultiplexers as multiplexers, which means that the ports once used as inputs in demultiplexers were designed for use as outputs according to the reciprocity principle, and vice versa, as shown in Figs. 5(a), (b), and (c). Structural analysis was performed using Lumerical FDTD for structures with a dielectric constant of $\varepsilon_{Si}$ (12.25), similar to that of continuous WDM structures [44]. The transmission efficiencies were similar to the WDMs. The results of our analysis show the use of structures as multiplexers for the first time in the literature, to the best of our knowledge. Because the structure is compact, reliable, and T-shaped, it can assist communication systems. These rudimentary results are also relevant to the burgeoning field of nanophononics.

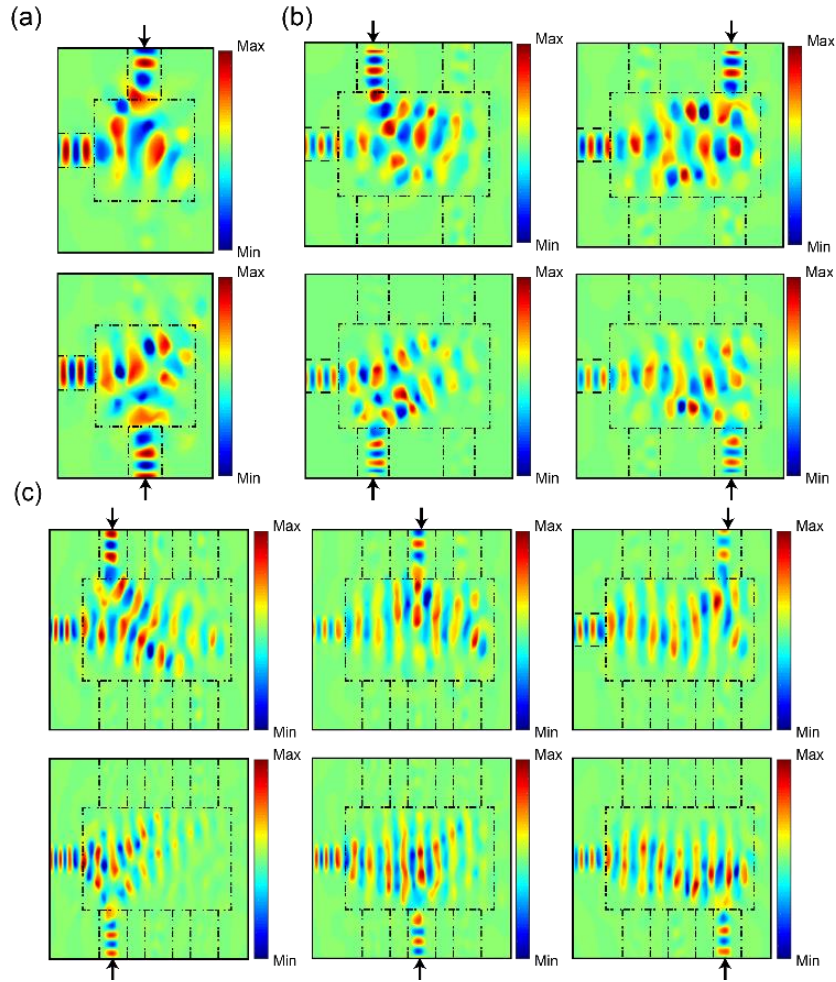

Fig. 5. H-field distributions of the continuous-dielectric (a) 1×2, (b) 1×4, and (c) 1×6 multiplexers. The arrows indicate the optical wave input to the ports.

## 4. CONCLUSION

This study proposed and numerically verified high-performance 1×N multiplexing systems. The resulting devices exhibited ultra-high transmission, small footprint, low crosstalk, and relatively narrow channel spacing, all of which are vital to a variety of applications of integrated photonics. Furthermore, the proposed all-dielectric 1×N T-junction (de)multiplexing systems implemented using the recently proposed inverse-design algorithm has significant potential for use in next-generation integrated photonics. Therefore, the proposed photonic designs are promising for next-generation on-chip photonic applications necessitating the superior properties of miniaturized optical components.

**ACKNOWLEDGMENTS**

We thank all members of Nanophotonics Lab of TOBB University of Economics and Technology. Hamza K. gratefully acknowledges the partial support of the Turkish Academy of Sciences.

Funding Information; Scientific and Technological Research Council of Turkey (TUBITAK) (116F200)